\documentstyle[12pt,epsfig]{article}
\def\beq{\begin{equation}}
\def\eeq{\end{equation}}
\def\be{\begin{equation}}
\def\ee{\end{equation}}
\def\bea{\begin{eqnarray}}
\def\eea{\end{eqnarray}}
\def\nnb{\nonumber}

\newcommand{\gsim}{\lower.7ex\hbox{$\;\stackrel{\textstyle>}{\sim}\;$}}
\newcommand{\lsim}{\lower.7ex\hbox{$\;\stackrel{\textstyle<}{\sim}\;$}}

\begin{document}

\begin{center}
\vspace{-3ex}{
                      \hfill hep-ph/0404042}\\[2mm]
{\LARGE\bf
Toward precision measurements in solar neutrinos} 

\vspace{0.6cm}
P. C. de Holanda$^a$, Wei Liao$^b$ and A. Yu. Smirnov$^{b,c}$\\ 
\vspace{0.3cm}
{\it $^a$ Instituto de F\'\i sica Gleb Wataghin - UNICAMP, 13083-970
Campinas SP, Brasil} \\
{\it $^b$ ICTP, Strada Costiera 11, 34014 Trieste, Italy} \\
{\it $^c$ Institute for Nuclear Research of Russian Academy of Sciences, 
Moscow 117312, Russia}
\end{center}
\begin{abstract}
Solar neutrino physics enters a stage of precision measurements. 
In this connection we present a precise analytic description of the 
neutrino conversion in the context of LMA MSW solution 
of the solar neutrino problem. Using the adiabatic perturbation theory 
we derive an analytic formula for the $\nu_e$ survival probability  
which takes into account the non-adiabatic corrections and the 
regeneration effect inside the Earth. The probability is averaged over 
the neutrino production region. We find that the non-adiabatic corrections 
are of the order $10^{-9}-10^{-7}$.   
Using the formula for the Earth regeneration effect 
we discuss features of the zenith angle dependence of the $\nu_e$ flux.  
In particular, we show that effects of small structures at the surface of 
the Earth can be important.

\end{abstract}

\section{Introduction}\label{sec1}

The LMA MSW solution~\cite{w,ms} has been identified
~\cite{sno1}$-$\cite{ped03} as
the correct solution of the solar neutrino problem. 
The $2 \nu $ conversion probability of this solution
gives a very good description of all available data: 
no statistically significant 
deviation has been found so far. 
New physics effects beyond LMA, if exist, are below few per cent. 

The program of future solar neutrino studies includes 

1). further tests of the LMA solution, in particular, searches for 
signatures of this solution such as the Day-Night asymmetry
and the distortion (``upturn") of the boron $\nu_e$
spectrum at low energies;

2). precise determination of the oscillation parameters, especially 
the 1-2 mixing angle;

3). searches for the sub-leading effects which originate from 

- 1-3 mixing, 

- sterile neutrino mixing, 
  
- non-standard neutrino interactions, 

- spin-flavor flip in the magnetic fields of the Sun, 

- violation of the fundamental symmetries 
(CPT, equivalence principle, {\it etc}.).\\

Already the present solar neutrino measurements have sensitivity 
at the level of few per cent.  
For instance, the predicted day-night asymmetry of the SuperKamiokande
signal is about $2\%$ which is comparable 
with the existing $1\sigma$ experimental error~\cite{sk-dn}. 
At SNO one expects the
$2-4\%$ asymmetry, consistent with the experimental
result~\cite{sno1} at the $1\sigma$ level.

Future experiments will have substantially higher 
sensitivity~\cite{BahPen,UNO,hyper-K,FREJUS}.
The solar neutrino studies enter a phase of 
precision measurements. \\

In this connection it is important 

\begin{itemize}

\item

to give precise description of the  LMA conversion, 
both in the Sun and in the Earth, taking into account various corrections; 

\item
to estimate accuracy of the approximations made; 

\item

to find  the precise {\it analytic} expressions for probabilities and 
observables 
as functions of the oscillation parameters 
($\Delta m^2$, $\sin^2 \theta_{12}$). 
This will help to test the LMA solution and to
search for physics beyond LMA.

\end{itemize}

We address these issues in the present paper.  
In section \ref{sec2} we consider the non-adiabatic corrections to the 
LMA conversion probability. We calculate these corrections for 
propagation inside the Sun and the Earth. In section \ref{sec3} we obtain
the analytical formula for the probability averaged 
over the distribution of
neutrino sources. In section \ref{sec4} we derive the analytic formula for 
the $\nu_e$ regeneration effect in the Earth. 
We present our conclusions in Section \ref{sec5}.
In the appendices A and B, alternative derivations of formulas
for the regeneration factor are given.

\section{Non-adiabatic corrections to the LMA solution}
\label{sec2}

According to the LMA MSW solution of the solar neutrino problem, a 
conversion of the solar electron neutrinos is driven by  
mixing of the two active neutrinos,
$\Psi_f \equiv (\nu_e, \nu_a)^T$ : 
\bea
\Psi_f = U(\theta) \Psi_{mass}, 
\eea 
where, in general, the mixing matrix is determined as 
\bea
U(\alpha) = 
\pmatrix{\cos\alpha & \sin \alpha \cr -\sin\alpha & \cos\alpha}, 
\label{matr}
\eea
and $\Psi_{mass} \equiv (\nu_1, \nu_2)^T$ is the vector of mass 
states.

\subsection{LMA and Adiabaticity}
\label{sec2.1}

The main feature of the LMA solution is the adiabaticity of 
conversion. According to LMA the averaged $2\nu$ 
survival probability of the 
electron neutrinos is given  
by the adiabatic formula~\cite{ms,messiah,parke}: 
\bea
P_{ee} = \frac{1}{2}(1+ \cos 2\theta^0_m \cos 2\theta).  
\label{adprob}
\eea
Here $\theta$ is the vacuum mixing angle, $\theta^0_m=\theta_m(x_0)$
is the mixing angle in matter in the neutrino
production point, $x_0$, and the mixing angle in matter
is determined by
\bea
\cos 2\theta_{m}(V) = \frac{\cos 2 \theta - 2 E V /\Delta m^2}
{[(\cos 2\theta - 2 E V /\Delta m^2)^2 + \sin^2 2\theta ]^{1/2}}, ~~
V=\sqrt{2} G_F n_e(x).
\label{cost}
\eea
Here $\Delta m^2$ is the mass squared difference, 
$E$ is the neutrino energy, $V$ is the potential,
$G_F$ is the Fermi coupling constant 
and $n_e(x)$ is the number density of electrons in the point $x$. 

How precise the expression (\ref{adprob}) is, 
and what are the non-adiabatic corrections? To answer these
questions, 
we will elaborate on the adiabatic perturbation theory.

Dynamics of neutrino conversion is described in terms of the 
instantaneous eigenstates of the Hamiltonian in matter, 
$\nu_{1m}$, $\nu_{2m}$. Representing an arbitrary neutrino state as 
$|\nu \rangle = \psi_{1m} |\nu_{1m} \rangle + \psi_{2m} | \nu_{2m} \rangle$, 
we can write the evolution equation in the base $(\nu_{1m}$, $\nu_{2m})$
as~\cite{ms,messiah,ms87}
\bea
i \frac{d}{dx} \pmatrix{\psi_{1m} \cr \psi_{2m}} =
\pmatrix{-\frac{\Delta(x)}{4 E} & -i {\dot \theta}_m(x) \cr
i {\dot \theta}_m(x) & \frac{\Delta(x)}{4 E} } 
\pmatrix{\psi_{1m} \cr \psi_{2m}},
\label{evoleigen}
\eea
where 
\bea
\Delta(x) \equiv   \Delta m^2 \sqrt{(\cos 2\theta-2 E V(x)/ \Delta m^2)^2 
+ \sin^2 2\theta} 
\label{massdiff}
\eea
and 
\be
{\dot \theta}_m(x) \equiv \frac{d \theta_m(x)}{dx}= 
\frac{E \Delta m^2 \sin2\theta}{\Delta(x)^2} \frac{d V(x)}{dx}.  
\label{tdot}
\ee

The adiabatic approximation corresponds to a situation when
\be 
\gamma \equiv \frac{4 E{|\dot \theta}_m|}{\Delta} \ll 1, 
\label{adcond}
\ee
and the off-diagonal terms in the Hamiltonian (\ref{evoleigen}) 
can be neglected. 
In this case there are no transitions between the eigenstates, 
and therefore the eigenstates propagate independently. The solution of 
(\ref{evoleigen}) is straightforward:
\be
\Psi_m^{ad}(x) = S^{ad} (x, x_0) \Psi_m (x_0),
\ee
where 
\be
\Psi_m^{ad}(x) = \pmatrix{ \psi_{1m}^{ad}(x) \cr
\psi_{2m}^{ad}(x)}, ~~~~
\Psi_m (x_0) = \pmatrix{ \psi_{1m}(x_0) \cr
\psi_{2m}(x_0)}, ~~~~
\label{defpsi}
\ee
and the adiabatic evolution matrix is
\be
S^{ad}(x, x_0) = S^{ad}(\Phi) = \pmatrix{e^{i \Phi (x)}& 0 \cr
0 & e^{-i \Phi(x)}}.
\label{evad}
\ee
The adiabatic phase $\Phi(x)$ equals
\be
\Phi(x) = \frac{1}{4E} \int_{x_0}^x  dx' \Delta (x'). 
\ee

A state initially produced as the electron neutrino,   
$\psi_{1m}(x_0) = \cos\theta_m^0$, 
$\psi_{2m}(x_0) = \sin \theta_m^0$, evolves as 
\be
\nu (x) = \cos\theta_m^0 e^{i \Phi(x)}\nu_{1m} + 
\sin \theta_m^0 e^{-i\Phi(x)} \nu_{2m}. 
\ee
The incoherent survival probability (\ref{adprob})  
can be immediately obtained 
by averaging the matrix element squared, $|\langle \nu_e |\nu 
(x)\rangle|^2$, 
over the oscillations.\\

\subsection{Non-adiabatic corrections}
\label{sec2.2}

The non-adiabatic corrections correspond to transitions between
the instantaneous eigenstates. We calculate these corrections
by solving the equation (\ref{evoleigen}).
We will implement a perturbation 
theory using the fact that for LMA the adiabaticity parameter 
(\ref{adcond}) is very small.

Let us search for the solution of the  
equation (\ref{evoleigen}) in the form
\bea
\pmatrix{\psi_{1m}(x) \cr \psi_{2m}(x)} && = 
\pmatrix{1 & c(x) \cr -c^*(x) & 1}
\pmatrix{\psi_{1m}^{ad}(x)  \cr \psi_{2m}^{ad}(x)} \nnb \\
&&=\pmatrix{e^{i \Phi (x)}& c(x)e^{-i \Phi(x)} \cr -c^*(x) e^{i \Phi(x)} & 
e^{-i \Phi(x)}}
\pmatrix{\psi_{1m}(x_0) \cr \psi_{2m}(x_0)},
\label{evolnew}
\eea
where $|c(x)| \ll 1$ is supposed to hold everywhere along 
the neutrino trajectory. 
(We check this {\it a posteriori}). 

The expression (\ref{evolnew}) can be rewritten as   
\be
\Psi_m (x) = S(x, x_0) \Psi_m (x_0), 
\ee
where the evolution matrix equals
\be
S(x, x_0) \equiv C S^{ad} \approx \pmatrix{e^{i \Phi (x)}& c(x)e^{-i 
\Phi(x)} \cr 
-c^*(x) e^{i \Phi(x)} &
e^{-i \Phi(x)}}. 
\label{ev}
\ee

Inserting  (\ref{evolnew}) into (\ref{evoleigen}) we find 
the differential equation for $c(x)$ 
from the condition that the off-diagonal elements 
of the evolution equation for  $\psi_{im}$ are zero: 
\be
i \frac{d}{dx}c(x)= -\frac{\Delta(x)}{2 E} c(x)-i {\dot \theta}_m(x).
\label{evolc}
\ee
Here the first order terms in $c(x)$ and ${\dot  \theta}_m$ are kept only. 
In this approximation 
the energy gap between the states coincides 
with the adiabatic split $\Delta(x)$ given in Eq. (\ref{massdiff}). 

The solution of equation (\ref{evolc}) can be written 
in the following form 
\bea
c(x) =- 
\int^x_{x_0} dx' \frac{d \theta_m(x')}{dx'}
\exp \left[-i \int^{x'}_{x} dx'' \frac{\Delta(x'')}{2 E}\right].
\label{newsolutb}
\eea
The integration constant is fixed  by the
condition : $c(x) \to 0$ as ${\dot \theta}_m \to 0$, so that  $c(x_0)=0$. 

Since for the LMA solution the phase $\Phi(x)$ is a fast oscillating 
function, 
the integral in (\ref{newsolutb}) can be calculated 
using the following formula (essentially the integration by parts)
\bea
\int^b_a f(x) e^{i g(x)} dx
= \left[ -i \frac{f(x)}{g'(x)}+\frac{f'(x)}{g'^2(x)}-
\frac{f(x) g''(x)}{g'^3(x)}\right] e^{i g(x)}\bigg|^b_a + {\cal O}(1/g'^3)
\label{approxm}
\eea
which is valid for smooth functions of $f(x)$ and $g(x)$. Here 
$g'(x)\equiv dg(x)/dx$ 
and $f'(x) \equiv df(x)/dx$. The formula  gives very good approximation if
$f(x)/g'(x) \ll 1$. In the case of integral (\ref{newsolutb})
this condition coincides with the adiabaticity condition
(\ref{adcond}) which is well satisfied. \\

According to (\ref{approxm}) and (\ref{newsolutb}) we find
\bea
c(x_f) && =-i \frac{2 E}{\Delta(x)}\frac{{d \theta}_m(x)}{dx}
\exp \left[-i \int^x_{x_f} dx' \frac{\Delta(x')}{2 E}\right]\bigg|^{x=x_f}_{x=x_0} \nnb \\
&& =-i {\rm sign}({\dot \theta}_m) \frac{\gamma(x)}{2}
\exp \left[- 2 i (\Phi(x)-\Phi(x_f)) \right] \bigg|^{x=x_f}_{x=x_0},
\label{csol1}
\eea
or explicitly
\bea
c(x_f)= - i \frac{2 E^2 \Delta m^2 \sin2\theta}
{\Delta(x)^3} \frac{d V(x)}{dx} \exp \left[-i \int^x_{x_f}
dx' \frac{\Delta(x')}{2 E}\right]\bigg|^{x=x_f}_{x=x_0},
\label{csol2}
\eea
where sign$({\dot \theta}_m)\equiv {\dot \theta}_m/|{\dot \theta}_m|$. 

We will apply this formula for neutrino propagation inside the Sun 
in section \ref{sec2.3} and inside the Earth in section \ref{sec4}.

\subsection{Non-adiabatic corrections for propagation inside the Sun}
\label{sec2.3}

The survival probability with the adiabaticity violation effect included 
can be written as  
\be
P_{ee} = \frac{1}{2}\left[1+ (1 - 2 P_c) \cos 2\theta^0_m \cos 
2\theta\right],
\label{surprob}
\ee
where $P_c=|c(x_f)|^2$ is the jump probability $-$ 
the probability of transition 
$\nu_{2m} \rightarrow \nu_{1m}$ on the way from $x_0$ to $x_f$. 

Let us calculate $P_c$. Notice that for the LMA solution, 
one can not use the Landau-Zener
formula~\cite{LanZen} for $P_c$ for the following reasons.
The mixing angle in the final point of evolution, $x_f$, is
large. The adiabaticity parameter $|4 E {\dot \theta}_m/\Delta|$
is of the same order for all points
inside $0.3$ of the solar radius~\cite{fried}.
The point of maximal adiabaticity violation is not the resonant
point, though not far from it. Moreover, the resonance layer defined as
$|2 E V-\Delta m^2 \cos 2\theta| \lsim \Delta m^2 \sin 2\theta$, is
broad since the vacuum mixing angle is large. Futhermore, significant part 
of the neutrino flux is
produced inside the resonance region or does not cross the resonance region    
at all. 

The double exponential formula~\cite{PetKra} is not valid too.  
It requires production of neutrinos far above the resonance region in the 
density scale. This formula is not applied in the range
\be 
\frac{\Delta m^2}{2 E}\cos 2\theta \sim 
(1.6 - 8.0)\cdot 10^{-6} ~~\frac{{\rm eV}^2}{\rm MeV},   
\label{range}
\ee
for which the density in the production point turns out to be
close to the resonance density. 
For the best fit values of the LMA oscillation parameters the range
(\ref{range}) corresponds to 
$E = (2 - 15)$ MeV, that is, to the region of interest. 

Let us apply the results of section 2.2 for calculation of the
non-adiabatic corrections.
Notice that at the surface of
the Sun the effective potential $V$ is negligible  and
${\dot \theta}_m$ can be taken zero. 
Using Eq. (\ref{csol2}) 
we find the transition amplitude $c(x_f)$ on the way from the 
production point to the surface of the Sun
in the leading order approximation as
\bea
c(x_f)= i \frac{2 E^2 \Delta m^2 \sin2\theta}
{\Delta(x)^3} \frac{d V(x)}{dx}\bigg|_{x=x_0} \times
\exp \left[i \int^{x_f}_{x_0} dx \frac{\Delta(x)}{2 E}\right].
\label{nonadinsunb}
\eea
Then the probability of  non-adiabatic transition 
is given by 
\be
P_c  =|c(x_f)|^2 = \frac{\gamma^2(x_0)}{4}
=\frac{1}{16 \pi^2} \frac{l^2_{osc}(x_0)}{h^2(x_0)}
\left[ \frac{2 E V(x_0) \Delta m^2 \sin 2\theta }{\Delta(x_0)^2} \right]^2, 
\label{nonadinsun}
\ee
where 
\be
h(x) \equiv  {V(x)}\left[\frac{d V(x)}{dx}\right]^{-1}, ~~~
l_{osc}(x) \equiv \frac{4 \pi E}{\Delta(x)}  
\ee
are the density height and the oscillation length in matter. 

The transition probability $P_c$ (\ref{nonadinsun}) 
depends only on parameters 
of the initial (production) point. 
One can understand this by noting that $l_{osc}(x)\ll h_c(x)$.
Therefore many oscillation lengths are obtained on the distance in which the
potential changes sizably. The 
non-adiabatic corrections 
are  averaged out being negligible along the trajectory of the neutrino 
except for the  boundaries of trajectory, {\it i.e.}, 
around the production point
or the point at the surface of the Sun. 
At the surface of the Sun the
contribution can be neglected because the potential is zero.

The probability is determined basically by  square of the  
ratio of the oscillation length and the density height. Second factor in 
(\ref{nonadinsun}) is of the order one. So, essentially $P_c \lsim
[l_{osc}(x_0)/4\pi h(x_0)]^2$.  
Using the best fit values of the LMA oscillation 
parameters we find from (\ref{nonadinsun})
\be
P_c = (10^{-9} - 10^{-7})\left(\frac{E}{10 ~{\rm MeV}}\right)^2. 
\label{pc}
\ee 
Here the numerical prefactor  depends on the production
point. As a function of $x_0$, the probability $P_c$ reaches maximum 
at around $ (0.1-0.2)R_{\odot}$,  
where the potential doesn't drop down substantially
and $h(x_0)$ reaches its almost minimal value due to increase of the 
density gradient. 
The corrections are negligible in the whole relevant range of neutrino
energies and production points. The probability (\ref{pc}) strongly differs
from what one would get using double-exponential formula~\cite{PetKra}:
$\sim e^{- 4\pi h /l_{osc}} \lsim 10^{-400}$.

Notice that the jump probability (\ref{nonadinsun}) equals (up to factor 4)
the adiabaticity parameter 
in the production point squared, as is expected in
the adiabatic perturbation theory. This contrasts the Landau-Zener
probability, $P_c \sim \exp(-\pi/2\gamma)$, which is essentially
non-perturbative effect.

\section{Averaging over production region: analytic results}
\label{sec3}

\begin{table}
\begin{center}
\begin{tabular}{|c|c|c|c|c|c|c|c|c|}
\hline
$K$ & pp & $^8B$ & $^{13}N$ & $^{15}O$ & $^{17}F$ & $^7Be$ & $pep$ & $hep$ 
\\
\hline
${\bar V}_K$($10^{-12}$eV) & 4.68 & 6.81 & 6.22 & 6.69 &6.74 & 6.16 & 5.13 & 3.96 
\\
\hline
$\Delta V_K^2/{\bar V}_K^2$ & 0.109 & 0.010 & 0.054 & 0.013 & 0.012 &
0.029& 0.076 & 0.165 \\
\hline
  \end{tabular}
 \end{center}
\caption{\it The average value of potential ${\bar V}_K$
and the corresponding value of
$\Delta V^2_K/{\bar V}_K^2$ for different component of the solar
neutrino spectrum.}
\label{tab:para}
\end{table}

In the adiabatic approximation the survival probability 
depends on the potential (density) in neutrino production point $r$: 
\be
P_{ee} = P_{ee}(V_0), ~~~ 
V_0=V(x_0=r).
\ee 

Observables at the Earth are determined by the survival probability 
averaged over the neutrino production region: 
\bea
P_K = \frac{\int dr ~G_K(r) P_{ee}(r)}{\int dr ~G_K(r)}, ~~~~
K = pp, pep, Be, N, O, F,  B, hep,  
\label{aveprob1}
\eea
where $G_K(r)$ is the distribution of 
sources of the $K$ component of neutrino spectrum.
The distributions 
are different for different components. 

Let us introduce the average value of the potential 
in the production region for the type $K$ neutrinos: 
\bea
\bar{V}_K \equiv \frac{\int dr ~G_K(r) V(r)}{\int dr ~G_K(r)}.
\label{avepot}
\eea
We will use the fact that in the effective production region, 
$V(r)$ deviates weakly from ${\bar V}_K$. Therefore the survival 
probability can be expanded in series around ${\bar V}_K$:
\bea
P_{ee} = P_{ee}(\bar{V}_K) + 
\left(\frac{d P_{ee}}{dV}\right)_{V = \bar{V}_K}(V - \bar{V}_K) + 
\frac{1}{2}\left(\frac{d^2 P_{ee}}{dV^2}\right)_{V = 
\bar{V}_K}(V - {\bar V}_K)^2 + \cdots
\eea
Inserting this expression into (\ref{aveprob1}) and using 
the definition (\ref{avepot}) we find 
\be
P_K  = P_{ee}(\bar{V}_K) 
-\frac{3 E^2}{(\Delta m^2)^2} \frac{\sin^2 2\theta \cos 2 \theta
(\cos 2 \theta - 2 E \bar{V}_K/\Delta m^2)} 
{[(\cos 2\theta- 2 E\bar{V}_K/\Delta m^2)^2+\sin^2 2\theta ]^{5/2}}
\Delta V^2_K,   
\label{aveprobb}
\ee
where 
\be
\Delta V^2_K \equiv
\frac{\int dr ~G_K(r) (V(r)- \bar{V}_K)^2}{\int dr ~G_K(r)}. 
\label{delpot2} 
\ee
Notice that the correction to $P_{ee}$ appears in the second order 
of the deviation of potential from the average value. 
The expression for probability (\ref{aveprobb}) can be rewritten as
\be
P_K = 
\frac{1}{2}+\frac{1}{2}( 1-\delta_K) \cos 2\theta_{m}(\bar{V}_K) \cos 
2\theta, 
\label{aveprob}
\ee
where the correction $\delta_K$ equals
\be
\delta_K = \frac{3}{2}\frac{(2 E \bar{V}_K/\Delta m^2)^2 \sin^2 2\theta}
{[(\cos 2\theta- 2 E\bar{V}_K/\Delta m^2)^2+\sin^2 2\theta ]^2}
\frac{\Delta V_K^2}{\bar{V}_K^2}.   
\label{deltaN}
\ee
In the Table \ref{tab:para} we present the average values of potentials 
and the corresponding second order deviations 
from the average values for eight types of solar neutrinos. We use 
the distributions of neutrino sources from the BP2000 model~\cite{bp2000}.
The expansion parameters $\Delta V_K^2/\bar{V}_K^2$ are all small,
especially for the boron neutrinos which have the narrowest
distribution of sources.

\begin{figure}[t]
\centerline{\psfig{figure=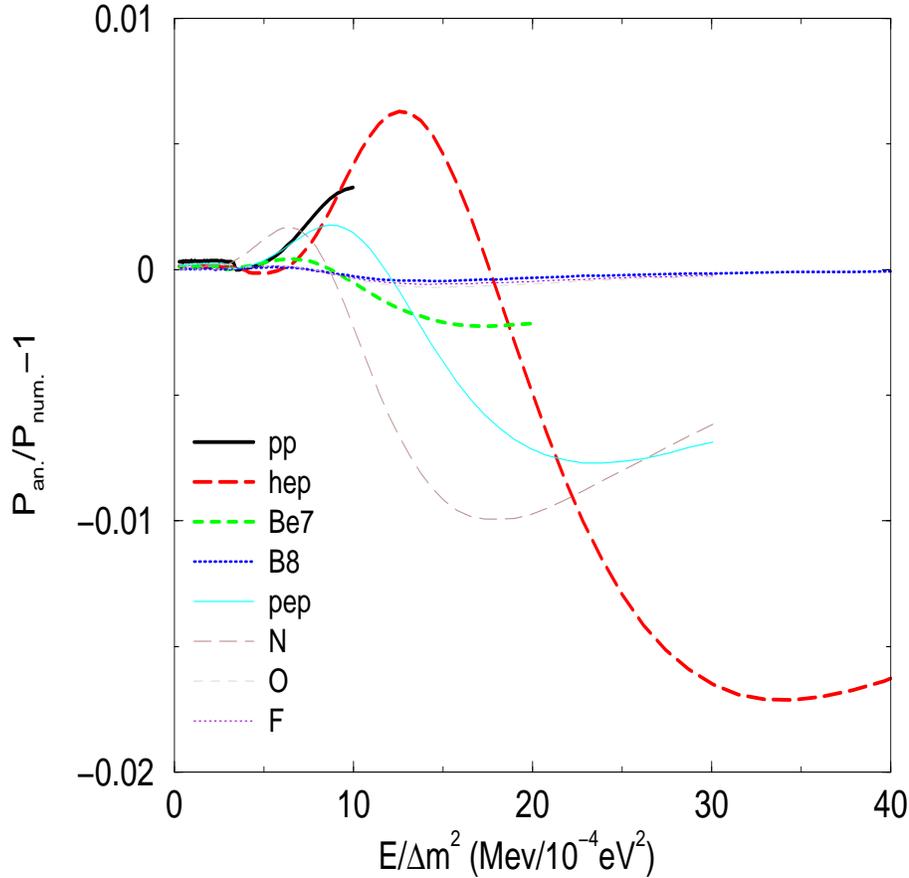,height=12cm,width=12cm}}
\caption{\small Deviations of the probability $P_K$ given in 
formula (\ref{aveprob}) from the
numerically calculated probability, $P'_K$ for different components of the 
solar neutrino spectrum.}
\label{figdiff}
\end{figure}

In Fig. \ref{figdiff} we compare the probability $P_K$ 
obtained from the approximate analytic formula  
(\ref{aveprob}) with results of numerical calculations, $P'_K$. 
$P_K/P'_K - 1$ is shown as a function of $E/\Delta m^2$.
The lines have been cut for $K=pp,Be7,pep,N,O,F$ because of their
lower energies in comparison with the  energies of $hep$ and boron 
neutrinos.
The plot shows that the analytic formula is rather precise.
In particular, the deviations are extremely small ($\lsim 10^{-3}$) for 
small and large 
values of $E/ \Delta m^2$. Relatively large deviations can be seen 
in the intermediate
region of $E/ \Delta m^2$. For example, for $K = hep$ the magnitude of
$P_K/P'_K-1$ reaches maximum $1.8\%$ at around
$E/\Delta m^2\approx 34 \times 10^{10}$~eV$^{-1}$.
The corrections $\delta_K$ are important: {\it e.g.}, for the
$hep$ neutrinos the 
deviation would be up to $6\%$ without $\delta_K$ .

\section{The Earth matter effect: analytic study}
\label{sec4}

The solar neutrinos arrive at the surface of the Earth as incoherent 
fluxes of the mass states. The mass states oscillate in the matter 
of the Earth producing partial regeneration of the
electron neutrino flux~\cite{MS86,DN,other,other1,other2,LMA,para_old,para}.  
Previously the effect has been described in one or two layers
approximation. In the later case, interference effects of
contributions from 
the core and the mantle have been discussed 
~\cite{para_old,para}. In this section we will
study effects for the realistic density profile
of the Earth.

\begin{figure}[t]
\centerline{\psfig{figure=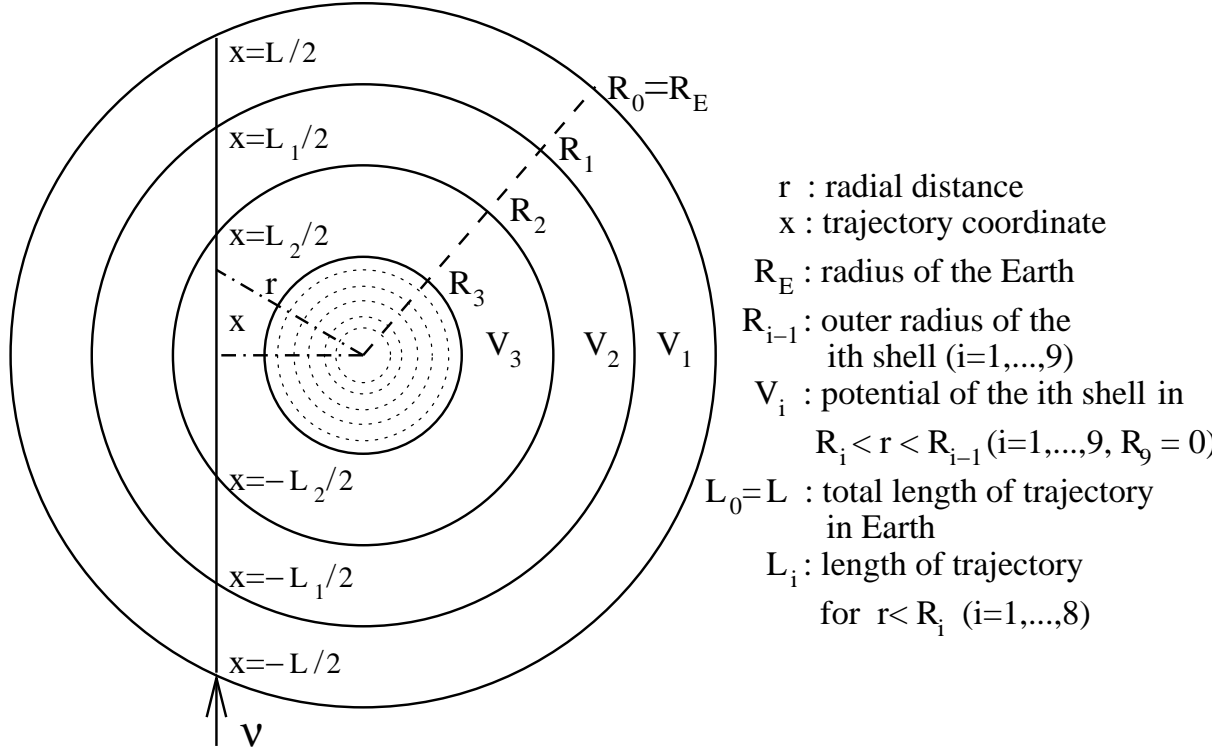,height=10cm,width=16cm}}
\caption{\small Structure of the Earth density profile. We indicate
notations used in the text.}
\label{figearth}
\end{figure}


\subsection{Regeneration factor and the Earth density profile}
\label{sec4.1}

The probability of $\nu_2 \to \nu_e$ transition can be 
written as 
\be
P(\nu_2 \to \nu_e) \equiv \sin^2\theta + f_{reg}, 
\label{freg}
\ee
where $f_{reg}$ is  the regeneration factor which  describes the 
Earth matter effect. 
In the absence of matter ({\it i.e.}, during the day) $f_{reg} = 0$. 
Using the definition (\ref{freg}) we find the
$\nu_e$ survival probability with the regeneration effect included 
as
\bea
P_{ee} = \frac{1}{2}(1+\cos 2\theta_m^0 \cos 2\theta) - 
\cos 2\theta_m^0 f_{reg}. 
\eea
Notice that the mixing angle $\theta_m^0$ in the neutrino production point 
in the Sun determines the mass ($\nu_1$, $\nu_2$) composition of the neutrino 
flux which arrives at the Earth. 

The essential feature of the LMA solution is that 
the Earth matter effect is small. This smallness 
is characterized by the ratio
\bea
\eta \equiv \frac{2 E V}{\Delta m^2} = 
0.024 \left(\frac{E}{10 {\textrm MeV}}\right) 
\left(\frac{6.3 \times 10^{-5} {\textrm eV}^2}{\Delta m^2}\right) 
\frac{V}{V_A}, 
~~~V_A=\sqrt{2} G_F N_A,
\eea
where $N_A$ is the Avogadro number. 
We will use $\eta$ as the expansion parameter.

The $\nu_2 \to \nu_e$ transition probability can be written as
\bea
P(\nu_2 \to \nu_e) && = |\langle\nu_e|U(\theta_{mR}) S(x_f,x_0)  
U^\dagger(\theta_{mR})U(\theta)|\nu_2\rangle|^2 \nnb \\
&& =  |\langle\nu_e|U(\theta_{mR}) 
S(x_f,x_0) U^\dagger(\theta_{mR} - \theta)|\nu_2\rangle|^2  ,
\label{p2e}
\eea
where $\theta_{mR}$ is the mixing angle in matter at the surface of the 
Earth and the matrix $S(x_f,x_0)$ given in
(\ref{ev}) describes evolution of the 
neutrino eigenstates in matter. 
Noting that in matter of the Earth 
\be
\sin(\theta_{mR} -\theta) \approx
\frac{E V_R}{\Delta m^2} \sin2\theta \ll 1,  
\ee
we find from (\ref{evolnew}), (\ref{freg}) and (\ref{p2e})
an expression for the regeneration 
factor in the lowest 
order in $c(x)$ and $\sin(\theta_{mR} -\theta)$ as
\bea
f_{reg} = \frac{2 E V_R}{\Delta m^2} \sin^2 2\theta  \sin^2 \Phi(x_f)
+ \sin 2 \theta  Re\{c(x_f)\},   
\label{regcor}
\eea
where $V_R$ is the potential at the surface of the Earth and
$\Phi(x_f)$ is the total phase acquired along the trajectory in the Earth.

For the profile with slowly changing density (lowest adiabatic approximation),
$c\approx 0$, we obtain
\bea
f^{ad}_{reg}=\frac{2 E V_R}{\Delta m^2}
\sin^2 2\theta \sin^2\Phi(x_f).
\label{regold}
\eea
The prefactor (the depth of oscillations) is determined
by the potential at the surface of the Earth, whereas the phase
is given by the integral along  whole trajectory.
For one layer with constant potential (density), and therefore 
$c = 0$, the regeneration factor (\ref{regcor}) or (\ref{regold}) 
is reduced to  
the well known expression: 
\bea
f_{reg}= \frac{2 E V_R}{\Delta m^2} \sin^2 2\theta 
\sin^2\frac{ \pi L}{l_m}.
\label{regold1}
\eea
Here $L$ is the distance traveled by neutrino in the Earth and $l_m$
is the oscillation length in matter. 

Let us consider a neutrino propagation in realistic density
profile of the Earth. The profile can be described by $n$ nearly
spherical shells of matter with sharp (step-like) density changes
at the borders of shells and slow variation of density
in layers between the borders. According to the PREM model 
$n=9$~\cite{PREM}.
So, in $i$th shell ($i=1,\cdots,n$), the potential $V_i$ is a smooth function
of the radial distance $r$. Crossing $j$ shells corresponds to
crossing $2j-1$ layers (see Fig. \ref{figearth}).
We denote by $R_{i-1}$ the outer radius of
$i$th shell, so that $R_0$ corresponds to the radius of the Earth: $R_0=R_E$.

The trajectory of the neutrino is 
characterized by the zenith angle $\theta_Z$. 
We determine a position of neutrino along trajectory by the 
coordinate $x$ 
with origin in the center of trajectory, so that
\bea
x \in [-L/2,L/2], ~~x^2=r^2-R_E^2\sin^2\theta_Z.
\label{trajcoor}
\eea
Here $L$ is the total length of the trajectory in the Earth.
The length of the part of trajectory inside border $R_i$
is given by
\bea
L_i = \sqrt{R_i^2-R_E^2\sin^2\theta_Z},
\label{length}
\eea
and $L_0=L$ by definition.

We introduce the adiabatic phase $\Phi_i$ acquired by neutrinos
in the interval $- L_i/2 \leq x \leq L_i/2$ along the trajectory,
that is, inside the outer border of the $(i+1)$th shell:
\bea
\Phi_i &&= \int^{L_i/2}_{-L_i/2} dx \frac{\Delta(x)}{4 E} \nnb \\
&&\approx \int^{L_i/2}_{-L_i/2} dx \bigg[ \frac{\Delta m^2}{4 E} - \frac{1}{2}
\cos2\theta V(x) + \frac{E \sin^2 2\theta}{2 \Delta m^2} V^2(x)\bigg] ,
~~~i=0,\cdots,n-1. 
\label{phasei}
\eea
$\Delta(x)$ is given in Eq. (\ref{massdiff}).
Here we keep the order $V^2$ term since due to integration its
contribution to the phase is not negligible.

At the borders of shells there are jumps of the potential 
and hence the discontinuities of the mixing angle in matter.
We denote them as
\bea
\Delta V_i && \equiv V_{i+1}(R_i)-V_i(R_i), 
~~i=0,\cdots,n-1, \label{jump1} \\
\Delta \theta_{mi} && \equiv \theta_m(V_{i+1}(R_i))- 
\theta_m(V_i(R_i)), ~~~
i=0, \cdots, n-1.
\label{jump}
\eea
At the surface of the Earth, we obtain
\bea
\Delta V_0= V_R, ~~~\Delta \theta_{m0} =\theta_{mR}-\theta.
\label{jump2}
\eea
Furthermore, we find
\bea
\sin \Delta \theta_{mi} \approx \frac{ E \Delta V_i}{\Delta m^2} \sin 2\theta, 
~~~\cos \Delta \theta_{mi} \approx 1, ~~~i=0,\cdots,n-1.
\label{approxth}
\eea
Corrections to (\ref{approxth}) are of the order $\eta^2$, and hence negligible.

Smooth variation of the potential in each shell of the Earth can be
approximated by the analytic formula~\cite{LisMon}:
\bea
V_i = V_A\left[ \alpha_i + \beta_i \frac{r^2}{R^2_E} +\gamma_i
\frac{r^4}{R^4_E}\right], ~~~i=1,\cdots,n.
\label{poteni}
\eea

Let us find an analytic expression for  the 
regeneration factor using the density profile described above.  
The problem can be solved in two steps: (1) computation of the  
non-adiabatic corrections to propagation within a given layer, $\Delta f_i$;
(2) computation of effect of the borders between layers, 
$\Delta f_i^{jump}$. So that
\bea
f_{reg} =f^{ad}_{reg} + \sum_i \Delta f_i+\sum_i \Delta f_i^{jump}.
\eea
The virtue of the LMA solution is that it enables us to study both 
effects using the same formalism of the adiabatic perturbation
theory.

\subsection{Non-adiabatic corrections in a layer of the Earth}
\label{sec4.2}

Let us compute the non-adiabatic corrections for one layer.  
Suppose a neutrino trajectory crosses the $i$th layer with the
borders at $x=L_i/2$ and $x=L_{i-1}/2$.
Using (\ref{trajcoor}) and (\ref{poteni}), the potential in
this layer can be expressed in terms of the trajectory coordinate as
\bea
V_i  = V_A\left[\alpha'_i+\beta'_i 
\frac{x^2}{R^2_E}+\gamma'_i\frac{x^4}{R^4_E}\right].  
\label{potenib}
\eea
Here
\bea
\alpha'_i=\alpha_i+ \beta_i \sin^2\theta_Z+\gamma_i \sin^4\theta_Z,~~
\beta'_i=\beta_i+2 \gamma_i \sin^2\theta_Z, ~~ \gamma'_i=\gamma_i.
\label{parametb}
\eea

According to (\ref{length}) and (\ref{parametb}) the gradients
of potential at the  borders equal 
\bea
\frac{d V_i(x)}{dx}\bigg|_{x={L_i \over 2}} &&
 = \frac{2 V_A}{R_E} \sqrt{y_i^2 - \sin^2\theta_Z}
(\beta_i+ 2 \gamma_i y_i^2),
\label{gradi} \\
\frac{d V_i(x)}{dx}\bigg|_{x={L_{i-1} \over 2}} &&=
\frac{2 V_A}{R_E} \sqrt{y_{i-1}^2 - \sin^2\theta_Z}
(\beta_i+ 2 \gamma_i y_{i-1}^2). 
\label{gradi-1}
\eea
where $y_i \equiv R_i/R_E$.
Then for this layer Eq. (\ref{csol2}) gives the 
amplitude of non-adiabatic transition
\bea
c_i&& =-2i e^{2 i \Phi_0} 
\frac{E^2 \Delta m^2 \sin 2\theta}{\Delta^3(x)} 
\frac{d V(x)}{dx} e^{-2 i \Phi(x)} 
\bigg|^{x={L_{i-1} \over 2}}_{x={L_i \over2}} \nnb \\
&& = -\frac{4 i E^2 \sin 2 \theta}{(\Delta m^2)^2 R_E} V_A e^{i \Phi_0}
\bigg[ (\beta_i+2 \gamma_i y_{i-1}^2 ) 
\sqrt{y_{i-1}^2 - \sin^2\theta_Z} 
e^{-i \Phi_{i-1}} \nnb \\
&& - (\beta_i+2 \gamma_i y_i^2)
\sqrt{y_i^2 - \sin^2\theta_Z} e^{-i \Phi_i} \bigg],
\label{ampli}
\eea
where phases $\Phi_i$ are defined in (\ref{phasei}).
Inserting this expression into (\ref{regcor}) we obtain the  non-adiabatic
correction from this layer, $\Delta f_{i}$ to the regeneration factor
as 
\bea
\Delta f_{i} &&= \frac{4 E^2 \sin^2 2 \theta}{(\Delta m^2)^2 R_E} V_A 
\bigg[ (\beta_i+2 \gamma_i y_{i-1}^2) \sqrt{y_{i-1}^2 - \sin^2\theta_Z}
\sin(\Phi_0 -\Phi_{i-1}) \nnb \\
&& - (\beta_i+2 \gamma_i y_i^2)\sqrt{y_i^2 - \sin^2\theta_Z}
\sin(\Phi_0 -\Phi_i) \bigg].
\eea

The ratio of the absolute value of correction and the adiabatic
term equals
\bea 
\frac{|\Delta f_i|}{|f^{ad}_{reg}|} \sim 
\frac{2 E }{\Delta m^2  \sin 2 \theta R_E} 
\frac{V_A (\beta_i + 2 \gamma_i y_i^2)}{V_R} 
\sqrt{y_i^2-\sin^2\theta_Z}
\sim \frac{2 E }{\Delta m^2}\frac{1}{R_E}, 
\label{ratio}
\eea
where $R_E$ plays the role of typical scale of the density change. 
As an example, let us consider
the layer between $0.895 R_E$ and $0.546R_E$. In this layer
$\alpha=3.156$, $\beta =-1.459$ and $\gamma=0.280$~\cite{LisMon}.
From (\ref{ratio}) we obtain that the non-adiabatic correction to the
regeneration factor is about $(1-2)\%$
at $E = 10$ MeV.

Notice that for some particular values of energies and $\theta_Z$,
the contributions $\Delta f_i$ from different layers $i$ may sum up
``constructively" producing larger effect. In this connection let us
notice the following.

1) The enhancement effect may occur for exceptional values of $E$ and
$\theta_Z$ and therefore any realistic averaging over $E$ and integration
over $\theta_Z$ will wash it out.

2) The enhancement can not be large (proportional to the number of
layers, $n$) since
(i) only few layers give significant contribution and for the rest,
the effect is below $1 \%$; (ii) there is a systematic cancellation
of contributions from the upper limit of integration in $\Delta f_i$
and the lower limit of integration in $\Delta f_{i-1}$ (the adiabatic
phases are the same for both); (iii) typically, contributions from
two layers of the same shell have opposite signs.

So, we conclude that the non-adiabaticity within
layers of the Earth can be safely neglected.

\subsection{Effects of several layers}
\label{sec4.3}

\begin{figure}[t]
\centerline{\psfig{figure=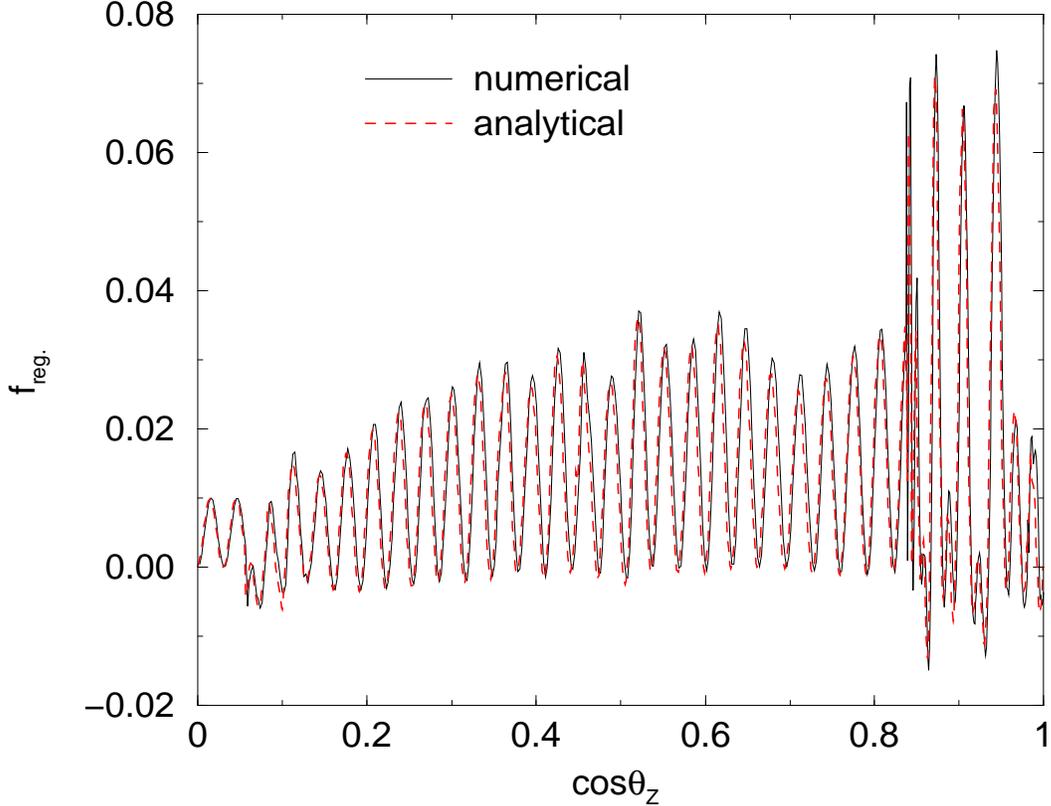,height=11cm,width=14cm}}
\caption{\small The regeneration factor as function of $\cos\theta_Z$ 
for $E=10$ MeV, $\Delta m^2=6.3 \times 10^{-5}$ ~eV$^2$, and 
$\tan^2\theta=0.4$. We compare result of numerical computations
with analytic result (\ref{regnewb}) for the PREM model. }
\label{figreg}
\end{figure}

The jumps of potential between the layers strongly violate the 
adiabaticity and on the first glance, the adiabatic
perturbation theory can not be applied. 
We show, however, that the results for non-adiabatic case 
obtained in section \ref{sec2.2} can be also used here.  
The key point is that for the LMA parameters  
the Earth matter effects are small,  whatever
the density profile in the Earth is. 
As a consequence, variations of the mixing angle in matter are small:
$|\Delta \theta_m| \ll \theta$, and essentially the expansion 
parameter here is $\eta$.  

Consider a neutrino trajectory which crosses 
$2 n-1$ layers ($n$ shells). In the points of the trajectory 
$x=\mp L_i/2$  neutrinos 
cross the borders of shell with  $r=R_i$, as illustrated 
in Fig. \ref{figearth}. The corresponding potential
jumps equal $\pm \Delta V_i$ for $x=\mp L_i/2$. 
Using the expression  (\ref{tdot}) we obtain in the lowest
approximation
\bea
{\dot \theta}_m(x) = \frac{ E \sin 2 \theta}{ \Delta m^2} 
\sum^{n-1}_{i=1} \Delta V_i 
\left[\delta \left(x+\frac{L_i}{2}\right) - 
\delta \left(x-\frac{L_i}{2}\right)\right].  
\label{theder}
\eea
As it has been shown in section \ref{sec4.2} to a good approximation 
one can take ${\dot \theta}_m = 0$ everywhere outside the borders.  

The evolution equation (\ref{evoleigen}) can be averaged in small 
intervals $\Delta x \ll 1/\Delta (x)$ 
to eliminate $\delta$-functions which originate from 
$\dot \theta_m$. 
However, this is not necessary since in the 
expression for $c(x)$ in (\ref{evolc}) $\dot \theta_m$ is integrated anyway.   

Plugging expression (\ref{theder}) into Eq. (\ref{newsolutb})
we obtain
\bea
c(L/2) 
 = \frac{ E \sin 2 \theta}{ \Delta m^2} e^{i \Phi_0 } 
\sum^{n-1}_{i=1} \Delta V_i \left(e^{-i \Phi_i} - e^{i \Phi_i}\right), 
\eea
where $\Phi_i$ are defined in (\ref{phasei}).

Inserting the real part of $c(L/2)$ into 
(\ref{regcor}) we find the regeneration factor for the case of
$n$ shells crossing:
\bea
f_{reg} =  \frac{2 E V_R}{\Delta m^2} \sin^2 2 \theta \sin^2 
\Phi_0
+ \sum^{n-1}_{i=1} \frac{2 E \Delta V_i}{\Delta m^2} \sin^2 2\theta
\sin\Phi_i \sin \Phi_0.
\label{regnew}
\eea
In the Appendix A we present the rigorous derivation
of this factor  considering evolution in the sequent layers explicitly. 
The results of two approaches coincide exactly in  the first 
order in $E V/\Delta m^2$. 

Noting that $V_R = \Delta V_0$ is the jump of potential at the 
surface of the Earth, we can rewrite the expression (\ref{regnew})
in the following compact form: 
\bea
f_{reg} = 
\frac{2 E \sin^2 2\theta}{\Delta m^2}\sin\Phi_0 
\sum^{n-1}_{i=0} \Delta V_i \sin\Phi_i.
\label{regnewb}
\eea
So, $f_{reg}$ is proportional to the sum of similar terms which
correspond to the borders of the shells. Each term is the product
of the potential jump at a given border and sine of the total adiabatic
phase acquired on the part of trajectory inside a given border (that is,
from $- L_i/2$ to $L_i/2$ for the border $i$). The sum runs over all
borders including the surface of the Earth. The expression (\ref{regnewb})
corresponds to the symmetric density profile.
The zenith angle dependence of the regeneration 
factor appears via the phases: 
$\Phi_i \equiv \Phi_i (\theta_Z)$.   

The formula (\ref{regnewb}) (which is the  main result of our study) 
allows us to get complete 
understanding of the Earth matter effects including effects 
of complicated  shell structure. 
Apparently, this is  not possible  using the one layer approximation 
(\ref{regold}), where the interference terms induced by different 
shells are absent.

In Fig. \ref{figreg} we compare the zenith angle dependence of the 
regeneration  
factor computed using the analytic formula (\ref{regnew}) with the one 
obtained by the exact numerical integration for the PREM profile. 
Two results coincide extremely well. One can see that the analytic
formula reproduces quite precisely the magnitude and the phase 
structure of the regeneration factor. 

Let us mark some features.
The change of the oscillatory behaviors for 
 $\cos\theta_Z \gsim 0.83$ is induced by the sharp density
jump at the border between the mantle and the core
of the Earth, at $r=0.54 R_E$. 
Notice that at $\cos\theta_Z \gsim 0.83$ the amplitude of oscillations
for some periods increases, however, the frequency of large peaks
becomes lower. So that the average value of $f_{reg}$ does not increase
in comparison with the  value for $\cos\theta_Z < 0.83$.

For small $\cos\theta_Z$ the dependence of $f_{reg}$ on 
$\theta_Z$ is a result
of interference of terms in (\ref{regnewb}) which correspond to 
the outer shells of the Earth. To understand this dependence, it is convenient
to introduce the phase $\varphi_i$:
\bea
\varphi_i \equiv \Phi_0 - \Phi_i,
\label{phased}
\eea
so that $\varphi_i/2$ is the phase acquired by neutrino on the way
from the surface of the Earth to $r=R_i$.
Using $\varphi_i$ we can rewrite the expression for
regeneration factor (\ref{regnewb}) as
\bea
f_{reg} =  \frac{2 E \sin^2 2\theta} {\Delta m^2}
\sum^{n-1}_{i=0} \Delta V_i
\left[\sin^2\Phi_0\cos\varphi_i - \frac{1}{2} \sin 2 \Phi_0 \sin\varphi_i
\right].
\label{newreg1}
\eea
Apparently, if averaging over $\varphi_i$ occurs only the
term with $i=0$ survives ($\varphi_0=0$) in $f_{reg}$ which is reduced to
the adiabatic expression for one layer.

The increase of regeneration factor
with $\cos\theta_Z$ in the range  $0.2 - 0.5$
is related to the effect of three density jumps 
near the surface of the Earth:  
according to the PREM profile they   
are situated  at depths 10 km, 22 km and 31 km correspondingly.
The distance, $L-L_i$, on  which $\varphi_i$ is acquired  
depends on the zenith angle as
\bea
L-L_i \approx 2 \frac{R_E-R_i}{\cos\theta_Z}, ~~
\cos\theta_Z > \sqrt{1-R_i^2/R^2_E}
\eea
($R_E-R_i$ is  the depth from the surface of the Earth to the borders
of the $i$th shell). 
In the case of small $\cos\theta_Z$ 
($\cos\theta_Z \lsim 0.2$), the distance 
$L-L_i$  can be of several hundreds kilometers which is comparable
to or larger than the oscillation length. Furthermore, $L-L_i$ and
$\varphi_i$ are fast changing functions of $\theta_Z$.
So, $\varphi_i$ are large and different for different $i$.
Therefore, the terms $\sin \Phi_0 \sin\Phi_i$ for different 
$i$ ($i=1,2,3$) are quite different and therefore partially cancel
each other (``interfere destructively'').

In constrast, for $\cos\theta_Z \gsim 0.5$,
the distances $L-L_i$ ($i=1,2,3$) for the outer shells become 
much smaller than 
the oscillation length and they slowly 
change with $\cos\theta_Z$. In this case the phases 
$\varphi_i$ ($i=1,2,3$) are all small and  
$\sin\Phi_0 \sin\Phi_i \approx \sin^2\Phi_0$.
So, for $\cos\theta_Z >  0.5$ the effects of outer shells
``interfere constructively'' producing larger regeneration factor.
It is then possible to account
these close layers effectively as a single layer, as it was
done in Ref.~\cite{LisMon}. 
Increase of the regeneration factor in the transition region
$\cos\theta_Z = (0.2-0.5)$ corresponds to convergence
of the term $\sin\Phi_0 \sin\Phi_i$ ($i=1,2,3$) to
$\sin^2\Phi_0$.

\subsection{Averaging over the neutrino energy}
\label{sec4.4}
\begin{figure}[t]
\centerline{\psfig{figure=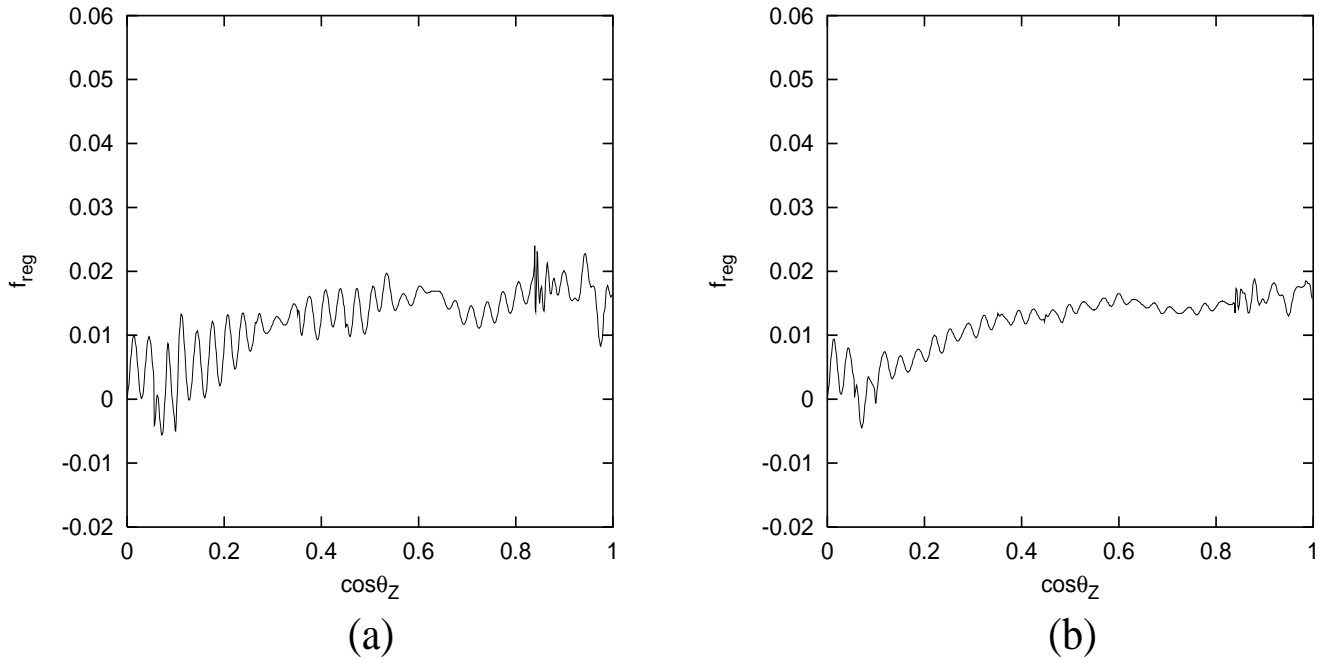,height=9cm,width=18cm}}
\caption{\small The regeneration factor averaged over the energy
 intervals (a) $E= (9.5 - 10.5)$ MeV; (b) 
$E= (8 - 11)$ MeV. For oscillation parameters we take 
$\Delta m^2=6.3 \times 10^{-5}$ ~eV$^2$ and
$\tan^2\theta=0.4$. }
\label{figave}
\end{figure}

For the LMA solution the oscillation length in the Earth is small:    
$l_m \approx l_{\nu} \ll R_E$.
Since the time of the neutrino detection is well known, 
averaging over the zenith angle can be avoided, 
and in fact, in the unbinned analysis of the data developed
recently~\cite{sk-dn} one needs to know the zenith angle dependence 
without averaging. 
At the same time since the recoil electron (and not 
neutrino) energy 
is measured and a detector has finite energy resolution,  
averaging over the neutrino energy occurs. 

In the leading approximation the phase equals $\varphi_i \approx
\Delta m^2 (L-L_i)/(4E)$. Therefore the energy resolution $\Delta E$
corresponds to averaging over the interval of phases:
\bea
\Delta \varphi_i \approx \varphi_i \frac{\Delta E}{E}.
\eea
If $L - L_i \gg l_m$, so that $\varphi_i \gg 1$,  
the interval $\Delta \varphi_i$ can
be large, thus leading to
strong averaging of terms $\sin\Phi_0 \sin\Phi_i$ in (\ref{regnewb}).
This happens to the contributions from  structures situated 
far from the surface of the Earth.
In Fig. \ref{figave} we show 
the result of averaging of the regeneration factor folded with  
the cross section of the neutrino-electron 
elastic scattering over two different energy intervals.  
Comparing with Fig. \ref{figreg}, one
sees that the complicated oscillatory pattern produced by the density jumps
in the central regions of the Earth 
is strongly averaged when $\cos\theta_Z$ 
is large (for general analysis of this effect see ~\cite{IS}). 
According to Fig. \ref{figave}
for $\cos\theta_Z \gsim 0.4$ the regeneration factor
oscillates with small depth around $f_{reg} \approx 1.5 \%$. This happens
because the main term $\sin^2 \Phi_0$ is strongly averaged too.

In contrast, for $\cos\theta_Z \lsim 0.2$ only the outer structures of the
Earth can contribute and the averaging is not as efficient
as for large $\cos\theta_Z$. Indeed, for the borders of outer
shells ($i=1,2,3$) the distance where $\varphi_i$ is acquired,
$L-L_i \le 2 R_E \sqrt{1-R_i^2/R^2_E}$, 
can be about several hundreds kilometers, that is, 
comparable with the oscillation length in matter. 
In this case the phase
interval $\Delta \varphi_i \approx \varphi_i\Delta E/E$
is still not large enough to give sufficient averaging.
For $\cos\theta_Z < 0.4$, $\Delta \Phi_0 \approx \Phi_0 \Delta E/E$ is 
also small and  averaging is weak.
Furthermore, in the interval $\cos\theta_Z
=(0.2-0.5)$ the regeneration factor increases with $\cos\theta_Z$.
The reason is that in this interval the main term, $\sin^2\Phi_0$,
starts to ``interfere constructively'' with the terms produced by the
outer shells, $\sin\Phi_0 \sin\Phi_i$ ($i=1,2,3$),
as it has been discussed in section \ref{sec4.3}.

\begin{figure}[t]
\centerline{\psfig{figure=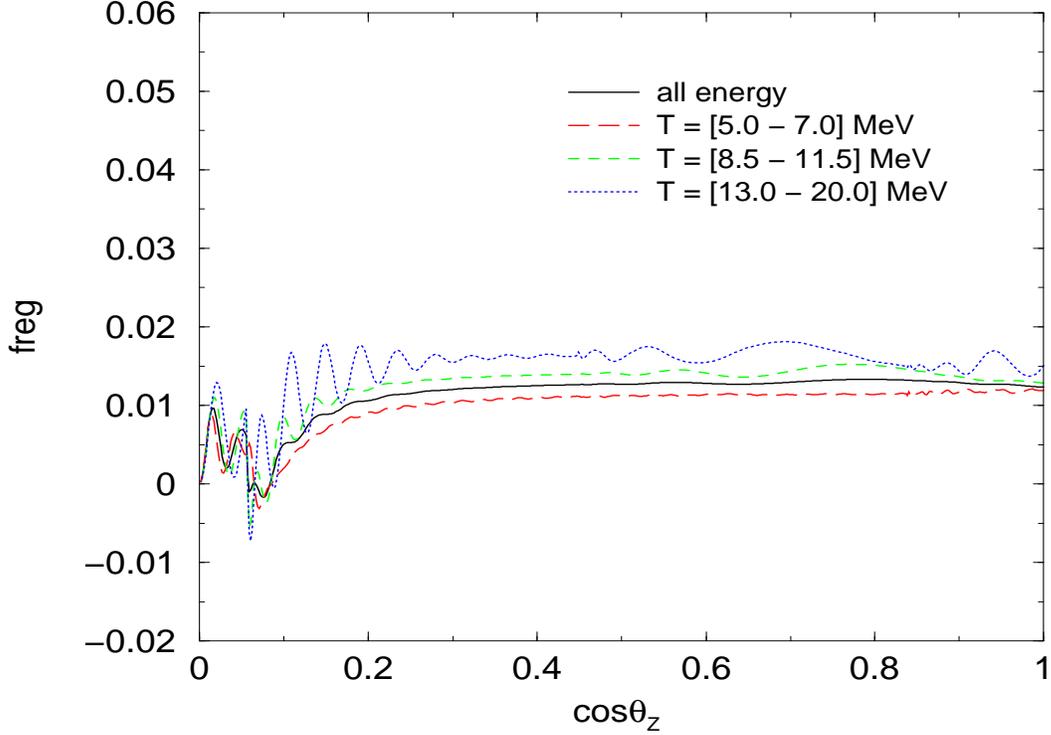,height=10cm,width=14cm}}
\caption{\small The regeneration factor for the SNO charged current
events integrated over  different intervals of the observed kinetic energy 
as function of the zenith angle. We take 
$\Delta m^2=6.3 \times 10^{-5}$ eV$^2$, $\tan^2\theta=0.4$.}
\label{snocc1}
\end{figure}

\begin{figure}[t]
\centerline{\psfig{figure=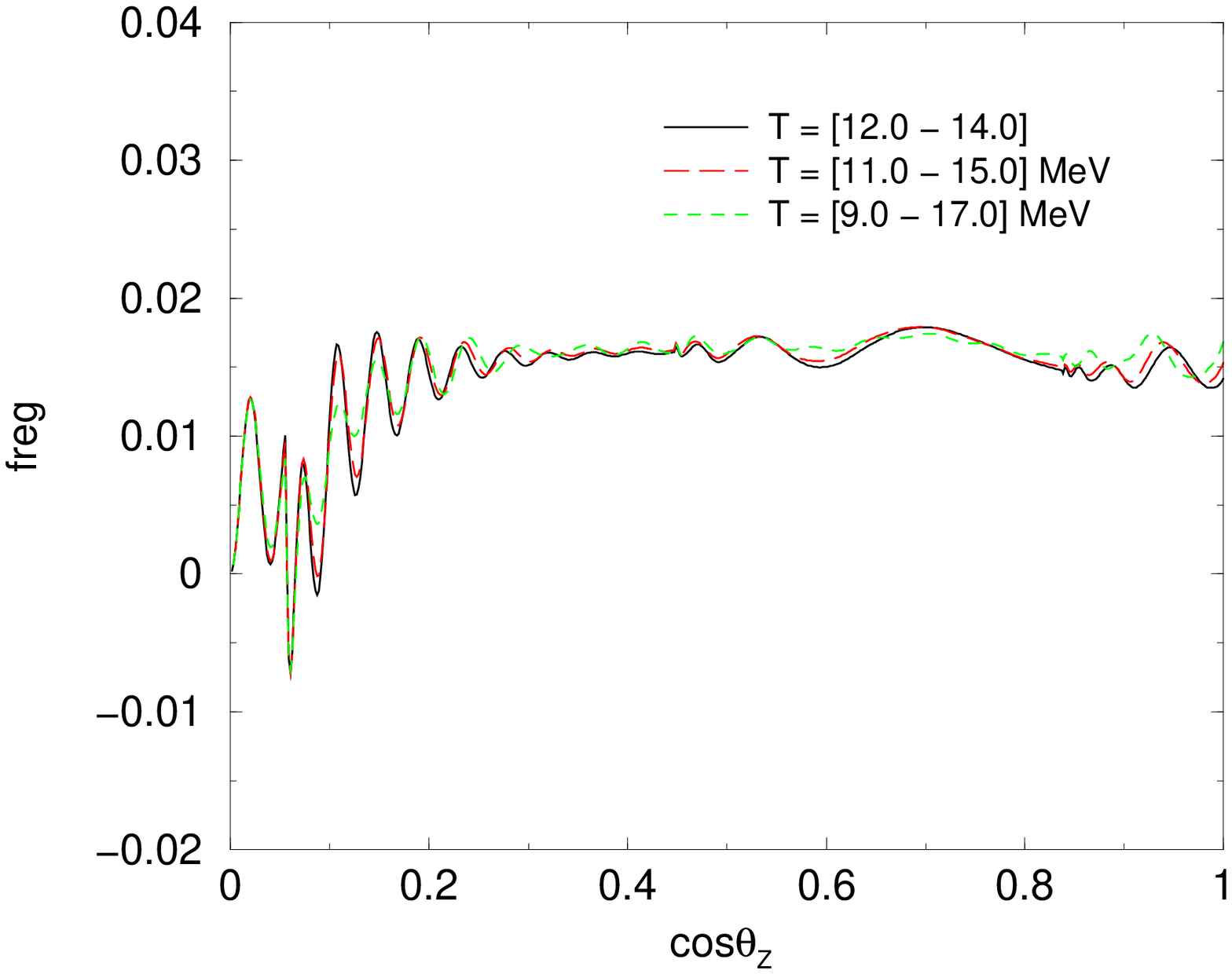,height=10cm,width=14cm}}
\caption{\small The same as in Fig. \ref{snocc1} for different intervals
of averaging and the same middle kinetic energy $T= 13$ MeV.}
\label{snocc2}
\end{figure}

In Fig. \ref{snocc1} and \ref{snocc2} we show  dependence of 
the regeneration factor
for the charged current events at SNO. Here we have taken into account
the energy resolution of the SNO detector and also performed
integration over various energy bins of the observed kinetic energy.
On the basis of our analytic formulas and discussion, the interpretation
of results of Fig. \ref{snocc1} and \ref{snocc2} is straightforward.

\subsection{Small scale structures: general density profile}
\label{sec4.5}

There are small scale structures  
in the  outer mantle of the Earth of
depth $(\sim 10)$ km in which matter has  
quite different densities ({\it e.g.}, ocean, rock and soil).  
In contrast to the ideal PREM 
model, these structures are not isotropically distributed and
can be quite complicated.

In section \ref{sec4.3} we have shown for the ideal PREM model
that contributions produced by structures close the surface of the Earth
interfere destructively for small $\cos\theta_Z$. 
Furthermore, averaging over the energy doesn't smooth
the dependence of these contributions on $\cos\theta_Z$ completely.
This produces an uncertainty for 
the future high statistics solar neutrino experiments (see also comments 
in ~\cite{LISI})
unless the local density distribution is well known~\cite{IS2}. 
In view of this we will consider 
general (not spherically symmetric) density profile.


Suppose neutrinos cross $k$ layers of matter. The density
jumps occur in the points
$x=x_i$ ($i=1,\cdots, k-1$);
$x=x_0$ and $x=x_k$ are the points where neutrinos enter and
leave the matter correspondingly.
Similarly to (\ref{theder}) we parameterize ${\dot \theta}_m$ as
\bea
{\dot \theta}_m= \frac{E \sin 2\theta}{\Delta m^2}
\sum^{k-1}_{i=1} \Delta V_i ~\delta(x-x_i),
\label{thederasy}
\eea
where the jump of potential at the $i$th border between layers equals
\bea
\Delta V_i =V(x_i+\epsilon)-V(x_i-\epsilon),  ~~~i=1,\cdots, k-1.
\label{jumpnewden}
\eea
$\epsilon$ is the infinitesimally small distance.
Noting that the potential is zero for neutrinos before entering 
the Earth and after leaving the Earth, we define also
\bea
\Delta V_0 = V(x_0),~~~\Delta V_k = -V(x_k).
\label{jumpnewden1}
\eea
Plugging the potential jumps into (\ref{newsolutb}) gives
\bea
c(x_k)= - \frac{E \sin 2\theta}{\Delta m^2}
\sum ^{k-1}_{i=1} \Delta V_i ~e^{2 i \phi_i},
\eea
where
\bea
\phi_i = \int^{x_k}_{x_i} dx \frac{\Delta(x)}{4 E}, ~~~i=0,\cdots, k
\label{phasex}
\eea
is the phase acquired from a given border $i$ to the final point
of the trajectory (detector).

Now it is straightforward to compute the regeneration factor, 
and in the leading order in $E V/\Delta m^2$ we obtain
\bea
f_{reg} &&= \bigg| \langle \nu_e| U(\theta_m(x_k)) S(x_k,x_0)
U^\dagger(\theta_m(x_0))U(\theta) | \nu_2 \rangle \bigg|^2 -\sin^2\theta \nnb \\
&& = -\frac{E \sin^2 2\theta}{\Delta m^2}
\sum^k_{i=0} \Delta V_i \cos 2\phi_i.
\label{regasy}
\eea
In the Appendix B, a direct computation of $f_{reg}$ for this case is given. 
Its result coincides with
(\ref{regasy}) in the leading order in $E V/\Delta m^2$.
Using this formula one can easily reproduce (\ref{regnewb})
assuming a symmetric density profile and taking into account
that $\Phi_i=(\phi_i-\phi_{k-i})$ for $k=2n-1$ and $i < n$.

Using (\ref{regasy}) it is easy to study 
averaging  effects following the discussion in  section \ref{sec4.4}. 
Structures situated far from the detector have $\phi_i \gg 1$. 
So that, after averaging, remote structures do not
produce significant effect. However, if the energy resolution
is improved, effects of these remote structures can be
large. This agrees with general consideration in ~\cite{IS}.

Thus, we arrive at the following conclusion.
If $\cos\theta_Z$ is large, small scale structures near
the entering point can be taken effectively as a single
layer. 
Furthermore, averaging over the energy makes contributions
of these small structures to be unimportant.

If $\cos\theta_Z$ is small, we can not consider small scale structures
near the entering point as a single layer. However uncertainties
produced by these structures can be significantly reduced if 
averaging is performed over broad energy interval, {\it i.e.} $\Delta E/E 
\sim 1$. 
After averaging, the regeneration factor still shows an oscillatory 
behavior in the region of small $\cos\theta_Z$, and this effect is  
produced by  contributions 
of the shells close to the detector~\cite{IS}. 

\section{Conclusion}
\label{sec5}

We have performed detailed analytic study of the
LMA MSW conversion of the solar neutrinos.
Our main result is the precise analytic formula for the survival 
probability which includes non-adiabatic corrections, averaging over
the neutrino production region and the Earth regeneration effect.
For the $K$ component of the solar neutrino spectrum
($K = pp, pep, Be, N, O, F,  B, hep$)
it can be written as
\bea
P_K = \frac{1}{2}+\frac{1}{2}(1-\delta_K) \cos 2\theta_m({\bar V}_K)
\cos 2\theta -(1-\delta_K) \cos 2\theta_m({\bar V}_K) f_{reg}.
\eea
Here the correction due to averaging effect, $\delta_K$, 
is given in Eq. (\ref{deltaN}); the average values
of matter potential in the production regions of $K$ components,
${\bar V}_K$, are defined in (\ref{avepot}) and their numerical
values are presented in the Table 1.
The regeneration factor $f_{reg}$ is given in (\ref{regnew})
for the symmetric density profile
and in (\ref{regasy}) for general asymmetric density profile.

Effect of averaging over the neutrino production region in the Sun
is reduced to specific value of the initial mixing angle in matter
which should be taken for the average value of the potential,
$\theta^0_m=\theta_m({\bar V}_K)$, and
to the appearance of the correction $\delta_K$.
We have compared the analytic results with the results of
numerical computation and found that maximal deviation $\sim 1.8 \%$
happens for the $hep$ neutrinos.
For the boron neutrinos the precision is better than 0.2\%.

We have obtained precise analytic formula
for the regeneration effect in the Earth using the
realistic  density profile.
We present simple derivation of this formula
which uses the adiabatic perturbation theory.
Performing also explicit calculations of the evolution
in sequent layers we show that this derivation is correct.
The analytic formula reproduces results of
numerical computations with accuracy determined by $\eta \sim 1-2 \%$.

Essentially the regeneration effect is the
sum of contributions from different shells which are determined by
jumps of the potential at the borders  and
by the adiabatic phase acquired inside the outer  borders of the
corresponding shells.
The dependence of regeneration factor on the zenith angle
can be understood in terms of interference of
contributions from different borders.

The derived analytical formula allows us to understand
the effect of averaging over the neutrino energy.
Using the analytical formula we have considered effects of
small scale structures ($\sim 10$ km) of the Earth profile.
These effects can be important for
small values of $\cos \theta_Z$.
We stress that local ``perturbations'' of the density profile
can produce sizable uncertainties in $f_{reg}$.\\ 

\noindent
{\Large \bf Acknowledgment}

One of the authors (P.C.H.) would like to thank FAPESP for financial
support. A.Y.S. thanks Tokyo Metropolitan University where this work 
has been accomplished for hospitality.  


\section*{\bf Appendix A. Regeneration factor in a symmetric density profile}\nonumber

Let us derive the  expression for the regeneration factor by considering
neutrino evolution in sequent layers of the Earth 
explicitly. 
First, we find  the complete evolution matrix, $\hat{S}$, in the
basis of the mass eigenstates $\Psi^T \equiv (\nu_1,\nu_2)$:
\bea
\Psi(x_f)=\hat{S}(x_f,x_0) \Psi(x_0).
\eea
As discussed
in section \ref{sec4.2}, the adiabaticity violation effect within
layers is suppressed by $2 E/(\Delta m^2 R_E) \sim 1-2\%$  
in comparison with the leading order Earth matter effect ($ \sim \eta$).
Therefore, we neglect the adiabaticity violation
within layers. 

1). In the case of neutrino propagation in one shell (one layer)
we can simply project the adiabatic evolution matrix (\ref{evad}) 
obtained for the matter eigenstates  
on to the basis of the mass states. 
In the leading order in $E V/\Delta m^2$ we find
\bea
{\hat S} && = \hat{S}_1 \bigg(\frac{L}{2},-\frac{L}{2}\bigg) \nnb \\ 
&& =U^\dagger (\theta) U(\theta_{mR}) S^{ad}\left(\frac{L}{2},-\frac{L}{2}\right)
U^\dagger (\theta_{mR}) U(\theta) =
U(\Delta \theta_{m0}) S^{ad}(\Phi_0)
U^\dagger(\Delta \theta_{m0}) \nnb \\
&&= S^{ad}\left(\frac{L}{2},-\frac{L}{2}\right) 
+ \sin\Delta \theta_{m0} (e^{-i \Phi_0}-e^{i \Phi_0})
\pmatrix{ \sin\Delta \theta_{m0} & \cos\Delta \theta_{m0} \cr
\cos\Delta \theta_{m0} & -\sin\Delta \theta_{m0} } \nnb \\
&& = \pmatrix{ e^{i \Phi_0} & 0 \cr 0 & e^{-i \Phi_0} }
+\frac{E \Delta V_0 \sin 2\theta}{\Delta m^2} (e^{-i \Phi_0}-e^{i \Phi_0})
\pmatrix{0 & 1 \cr 1 & 0 }.
\label{matr1}
\eea
Here $\Delta \theta_{m0} \equiv \theta_{mR} - \theta$ is the jump of the mixing 
angle at the surface of the Earth. We have used Eq. (\ref{approxth}), and 
$\Phi_0$ is defined in Eq. (\ref{phasei}).

2).  In the case of two shells crossing, the neutrino encounters 
three layers (the outer shell is crossed twice). The evolution
matrix can be similarly obtained by using the adiabatic evolution
matrix (\ref{evad}) in each layer and  by  rotation from the 
matter eigenstates basis in the layer before the border 
to the basis after the border.   
As a result, we find
\bea
\hat{S} = U(\Delta \theta_{m0}) 
S^{ad}\left(\frac{L}{2},\frac{L_1}{2}\right) 
S_1 \bigg(\frac{L_1}{2},-\frac{L_1}{2} \bigg) 
S^{ad}\left(-\frac{L_1}{2},-\frac{L}{2}\right) 
U^\dagger(\Delta \theta_{m0}), \nnb
\eea
where $S_1$ is the evolution matrix in the inner shell which has
a form similar to Eq. (\ref{matr1}) and it can be written as 
\bea
S_1 && = U(\Delta \theta_{m1}) S^{ad}\left(\frac{L_1}{2},-\frac{L_1}{2}\right)
U^\dagger(\Delta \theta_{m1}) \nnb \\
&& = S^{ad}\left(\frac{L_1}{2},-\frac{L_1}{2}\right)
+ \frac{E \Delta V_1 }{\Delta m^2} \sin 2\theta
(e^{-i \Phi_1}-e^{i \Phi_1})
\pmatrix{ 0 & 1 \cr
1 & 0 }. 
\label{matr2b}
\eea
Here $\Delta \theta_{m1}$ is the jump of mixing angle on the border
between the first and the second shells, and  $\Phi_1$ is defined
in Eq. (\ref{phasei}).

Combining the last two formulas we find to the order $E V/\Delta m^2$
\bea
{\hat S}  &&=  {\hat S}_2\left(\frac{L}{2},-\frac{L}{2}\right) \nnb \\
&& =S^{ad}\left(\frac{L}{2},-\frac{L}{2}\right)
+\sum_{i=0}^1 \frac{E \Delta V_i \sin 2\theta}{\Delta m^2} 
(e^{-i \Phi_i}-e^{i \Phi_i})
\pmatrix{0 & 1 \cr 1 & 0 } .
\label{matr2}
\eea

3). Suppose the evolution matrix for $j$ shells crossing ($2j-1$ 
layers) equals 
\bea
{\hat S}  &&=  {\hat S}_j \left(\frac{L}{2},-\frac{L}{2}\right) \nnb \\
&& =S^{ad}\left(\frac{L}{2},-\frac{L}{2}\right)
+\sum_{i=0}^{j-1} \frac{E \Delta V_i \sin 2\theta}{\Delta m^2} 
(e^{-i \Phi_i}-e^{i \Phi_i})
\pmatrix{0 & 1 \cr 1 & 0 }. 
\label{matrj}
\eea
Consider now the trajectory with $j+1$ shells crossings. The evolution
matrix is
\bea
{\hat S}&&= {\hat S}_{j+1} \nnb \\
        &&=\prod^{j-1}_{i=0} \bigg[ U(\Delta \theta_{mi}) 
S^{ad}\left(\frac{L_i}{2},
\frac{L_{i+1}}{2}\right) \bigg] S_j \bigg(\frac{L_j}{2},-\frac{L_j}{2}\bigg)
\prod^{0}_{i=j-1} \bigg[ S^{ad}\left(-\frac{L_{i+1}}{2},-\frac{L_i}{2}\right)
U^\dagger(\Delta \theta_{mi}) \bigg], 
\label{sss}
\eea
where $\Delta \theta_{mi}$ defined in (\ref{jump}), is the jump of the mixing 
angle in matter at the border $R_i$.
$S_j$ is the evolution matrix in the central shell which can be 
written similarly to Eq. (\ref{matr2b}) as 
\bea
S_j
 = S^{ad}\left(\frac{L_j}{2},-\frac{L_j}{2}\right)
 + \frac{E \Delta V_j}{\Delta m^2} \sin 2\theta
(e^{-i \Phi_j}-e^{i \Phi_j}) \pmatrix{ 0 & 1 \cr 1 & 0 }
\label{matrjb}.
\eea
$\Phi_j$ is given in Eq. (\ref{phasei}). 
After insertion into (\ref{sss}), the first 
term of (\ref{matrjb}) leads to ${\hat S}_j$.
The second term in (\ref{matrjb}) is already of the 
order $E V/\Delta m^2$. Note that $\Delta \theta_{mi}$ is small, 
as is shown in (\ref{approxth}). So,  we can 
approximate $U(\Delta \theta_{mi})$ by the  unit matrix when
plugging the second term in (\ref{matrjb}) into (\ref{sss}).
As a result, we find
\bea
{\hat S} = {\hat S}_j +\frac{E \Delta V_j \sin 2\theta}{\Delta m^2}
(e^{-i \Phi_j}-e^{i \Phi_j}) \pmatrix{0 & 1 \cr 1 & 0 }.
\eea
Using then expression (\ref{matrj}) for ${\hat S}_j$, 
the formula (\ref{matrj}) is immediately extended to the
case of crossing $j+1$ shells, thus accomplishing the proof. 
The result for the case of $n$ shells crossing 
is 
\bea
{\hat S}  
=S^{ad}\left(\frac{L}{2},-\frac{L}{2}\right)
+\sum_{i=0}^{n-1} \frac{E \Delta V_i \sin 2\theta}{\Delta m^2}
(e^{-i \Phi_i}-e^{i \Phi_i})
\pmatrix{0 & 1 \cr 1 & 0 } .
\label{matrn}
\eea

Using (\ref{matrn}) we obtain the expression for the regeneration
factor in the leading order in $E V/\Delta m^2$ as 
\bea
f_{reg} && =\bigg| \sin\theta e^{-i \Phi_0}+\cos\theta \sum_{i=0}^{n-1}
\frac{E \Delta V_i \sin 2\theta}{\Delta m^2}
(e^{-i \Phi_i}-e^{i \Phi_i}) \bigg|^2 -\sin^2\theta \nnb \\
&&= \frac{2 E \sin^2 2\theta}{\Delta m^2} \sin\Phi_0
\sum_{i=0}^{n-1}\Delta V_i \sin\Phi_i,
\label{regen}
\eea
where $\Phi_i$ is given in (\ref{phasei}) and $\Delta V_i$ is defined
in (\ref{jump1}).
This expression coincides with (\ref{regnew}) or (\ref{regnewb})
which have been obtained in section \ref{sec4.3} using 
the adiabatic perturbation
theory.

\section*{\bf Appendix B. Regeneration in asymmetric density profile}
\nonumber

As in the section \ref{sec4.5}, we define
\bea
\Delta \theta_{i} \equiv \theta_m(x_i+\epsilon)-\theta_m(x_i-\epsilon),
~~~i=0,\cdots, k,
\eea
where $\Delta \theta_{0}=\theta_m(x_0)-\theta$ and 
$\Delta \theta_{k}= \theta - \theta_m(x_k)$, and $x=x_i$
are the points of density jumps.
$x_0$ and $x_k$ are the initial and final points of neutrino trajectory 
in matter.
We will use the following expression:
\bea
\sin \Delta \theta_{i} \approx \frac{E \Delta V_i}{\Delta m^2} \sin 2\theta,
~~~\cos\Delta \theta_{i} \approx 1,
\label{approxb}
\eea
which is a good approximation in the leading order in $E \Delta V/\Delta m^2$.
$\Delta V_i$ is given in (\ref{jumpnewden}).
Neglecting the adiabaticity violation within each layer,
we obtain the evolution matrix ${\hat S}$ as 
\bea
{\hat S} = \bigg (\prod^1_{i=k} U^\dagger(\Delta \theta_{i})
S^{ad}(x_i,x_{i-1}) \bigg) U^\dagger(\Delta \theta_{0}).
\label{matrasy}
\eea
$S^{ad}(x,x_0)=S^{ad}(\phi(x))$ is the adiabatic
evolution matrix given in (\ref{evad}).

We approximate $U^\dagger(\Delta \theta_{i})$ as
\bea
U^\dagger(\Delta \theta_{i}) = \pmatrix{1 & 0 \cr 0 & 1}+
Q(\Delta \theta_{i}),
\eea
where
\bea
Q(\Delta \theta_{i}) \equiv \pmatrix{0 & -\sin\Delta \theta_{i} 
\cr \sin \Delta \theta_{i} & 0}.
\eea
Straightforward computation gives the following result in the leading order in 
$E V/\Delta m^2$ 
\bea
{\hat S} 
&&= S^{ad}(x_k,x_0)+Q(\Delta \theta_{k}) S^{ad}(x_k,x_0)
+S^{ad}(x_k,x_0) Q(\Delta \theta_{0}) \nnb \\
&& +\sum^{k-1}_{i=1} S^{ad}(x_k,x_i) Q(\Delta \theta_{i}) 
S^{ad}(x_i,x_0)\nnb \\
&&  = S^{ad}(x_k,x_0)+ \sum^k_{i=0} 
\pmatrix{0 & -\sin\Delta \theta_{i} e^{i(2 \phi_i-\phi_0)} \cr
 \sin \Delta\theta_{i} e^{-i(2 \phi_i-\phi_0)} & 0}.
\eea
$\phi_i$ is defined in (\ref{phasex}). Then,  
using (\ref{approxb}), the regeneration factor can be  directly 
computed in the first order in $E V/\Delta m^2$ as
\bea
f_{reg} &&= | \sin\theta e^{-i \phi_0}-\cos\theta \sum^k_{i=0}\
\sin\Delta \theta_{i} e^{i(2 \phi_i-\phi_0)} \bigg|^2 -\sin^2\theta \nnb \\
&& = - \frac{E \sin^2 2 \theta}{\Delta m^2} 
\sum^k_{i=0} \Delta V_i \cos 2 \phi_i.
\eea
It coincides with (\ref{regasy}).
\\

\noindent
{\Large \bf Note added}\\

This note has been added on request of the referee.

1). After the present paper had appeared in the hep-ph archive
[hep-ph/0404042], the preprint by Akhmedov {\it et. al.},
[hep-ph/0404083],
has been published in which the analytic integral formula is given for the
the regeneration effect in the Earth in the three
neutrino framework. In the first version of
[hep-ph/0404083] the correct oscillation phase in this integral formula
has been introduced on the ``heuristic'' basis: it does not follow
from their perturbation theory.
Correct integral formula (with the correct phase) has been derived for the
first time  in the paper
by Ioannisian and  Smirnov, [hep-ph/0404060].
Later in the Journal version JHEP 0405 (2004) 057,
Akhmedov {\it et. al.}, have also presented derivation of correct phase.

Let us now compare the results of papers [hep-ph/0404060], 
[hep-ph/0404083] with the results of present paper.
(We will use the Akhmedov's {\it et. al.} results in the
limit of zero 1-3 mixing.)

In Ioannisian and Smirnov paper hep-ph/0404060 and Akhmedov {\it et. al.}
paper JHEP 0405 (2004) 057 the integral formula has been obtained
using the improved perturbation theory in the small parameter
$\eta \equiv 2E V(x)/\Delta m^2$.
In the present paper we use the adiabatic perturbation theory.
It can be  shown that in the lowest order in  $\eta$
both approaches coincide. Indeed, inserting expression for
${\dot \theta}_m$ from (\ref{tdot}) into (\ref{newsolutb}) and
performing integration by parts in Eq. (\ref{newsolutb}) of the present paper
one can derive the integral formula.

This formula is valid for arbitrary density profile provided 
that the condition $\eta \ll 1$ is satisfied. Inserting the potential
of PREM model (Eqs. (\ref{jump1}, \ref{potenib})) into the integral formula
one can reproduce the result (\ref{regnewb}). 
However technically the use of
formula (\ref{newsolutb}) of the present paper is more convenient 
for the derivation of
(\ref{regnewb}) since the derivative $d\theta_m/dx$ gives $\delta$- functions
at the borders of layers and further integration becomes trivial.

2). The KamLAND collaboration has published recently
results of improved measurements of oscillations on the
basis of 766.3 ton-year exposure [T. Araki {\it et. al.}, hep-ex/0406035].
In assumption of the CPT conservation,
the global analysis of the solar neutrino data and KamLAND
gives slightly ($\sim 10\%$) higher best fit value $\Delta m^2 = 8 \cdot
10^{-5}$ eV$^2$ than it was before.
The increase of $\Delta m^2$ leads to the corresponding small decrease of
the adiabaticity parameter $\gamma$ and the expansion
parameter $\eta$ for a given energy
of neutrinos. Therefore increase in $\Delta m^2$ (i) further improves
the adiabatic perturbation theory and
implies that the non-adiabatic correction for probability in the Sun
is smaller;  (ii) diminishes the Earth matter regeneration effect;
(iii) makes our analytic formula for $f_{reg}$ 
preciser. Notice that the  analytic study of this paper has a general
character and does not rely on particular values of $\Delta m^2$.
We use specific  value of $\Delta m^2$ for illustration only.

\end{document}